\documentclass{article}

\usepackage{graphicx,hyperref}

\usepackage{amsthm,amsfonts,amsopn,amsmath,amssymb,natbib}
\usepackage[latin1]{inputenc}
\usepackage{xspace}
\usepackage{float}
\usepackage{algorithm,algorithmic,color}
\usepackage[section]{placeins}
\usepackage{sidecap}
\usepackage[top=1.93in, bottom=1.6in, left=1.6in, right=1.6in]{geometry}
\usepackage{microtype}

\DeclareMathOperator{\Var}{Var}

\DeclareMathOperator{\D}{D}

\DeclareMathOperator{\Cov}{Cov}
\DeclareMathOperator{\Id}{I}
\newcommand{\npar}{\par \vspace{2.3ex plus 0.3ex minus 0.3ex}}
\newcommand{\E}{\mathbb{E}}

\theoremstyle{plain}

\def\E{\mathbb{E}} 

\author{Tim Salimans\footnote{\href{mailto:salimanstim@gmail.com}{salimanstim@gmail.com}}}
 
\title{Implementing and Automating\\ Fixed-Form Variational Posterior Approximation\\ through Stochastic Linear Regression}

\begin{document}

\renewcommand{\thefootnote}{\fnsymbol{footnote}}

\maketitle

\renewcommand{\thefootnote}{\arabic{footnote}}

\begin{abstract} We recently proposed a general algorithm for approximating nonstandard Bayesian posterior distributions by minimization of their Kullback-Leibler divergence with respect to a more convenient approximating distribution. In this note we offer details on how to efficiently implement this algorithm in practice. We also suggest default choices for the form of the posterior approximation, the number of iterations, the step size, and other user choices. By using these defaults it becomes possible to construct good posterior approximations for hierarchical models completely automatically. 
\end{abstract}

\section{Introduction}
\label{intro}
In Bayesian analysis our main goal is to obtain and analyse posterior distributions of the form $p(x|y)$, where $x$ contains unknown parameters and latent variables that we would like to infer from observed data $y$. This is diffult because the form of the posterior distribution $p(x|y)$ is often not analytically tractable. To obtain quantities of interest under such a distribution, such as moments or marginal distributions, we typically need to use Monte Carlo methods or approximate the posterior with a more convenient distribution. A popular method of obtaining such an approximation is \textit{structured} or \textit{fixed-form Variational Bayes}, which works by numerically minimizing the Kullback-Leibler divergence of a parameterized approximating distribution $q_{\eta}(x)$ of more convenient form to the intractable target distribution~\citep{Attias2000,Beal2006,jordan1999introduction,wainwright2008graphical}.

\begin{equation}
\label{eq:vbapprox}
\hat{q} = \arg\min_{q(x)} \D[q|p] = \arg\min_{q(x)} \E_{q(x)}\left[\log \frac{q(x)}{p(x,y)}\right], 
\end{equation}

where $p(x,y)$ denotes the unnormalized posterior distribution $p(x|y)p(y)$. 
\npar
Standard methods for solving \eqref{eq:vbapprox} assume that the expectation $\E_{q(x)}$ in this expression can be calculated analytically, which is only the case for a very limited combination of approximations and posterior distributions. In contrast, our approach is based on Monte Carlo approximation, and is therefore applicable to a much wider range of applications. To be precise, we only assume that we can evaluate $\log p(x,y)$ up to a normalizing constant, and we allow for a more general specification of the approximation $q(x)$ of the form

\begin{equation}
q_{\eta}(x) = \prod_{i} q_{\eta_{i}}(x_{i}|x_{p_{i}}),
\end{equation}

with one or more (possibly multivariate) blocks of unkowns $x_{i}$, and where $x_{p_{i}} = x_{1},x_{2},\ldots,x_{i-1}$ denotes the parents of $x_{i}$, i.e. the unknowns higher up in the hierarchy of $q_{\eta}(x)$. We further assume that each conditional $q_{\eta_{i}}(x_{i}|x_{p_{i}})$ is a tractable member of the exponential family. That is,

\begin{equation}
\label{eq:expfam}
\log q_{\eta_{i}}(x_{i}|x_{p_{i}}) = T(x_{i}; x_{p_{i}})\eta - Z(\eta,x_{p_{i}}),
\end{equation}

with $T(x_{i}; x_{p_{i}})$ a $1 \times k$ vector of sufficient statistics, and $Z(\eta,x_{p_{i}})$ a normalizing constant.
\npar
In earlier work \citep{linregvb} we showed that solving the optimization problem of fixed-form Variational Bayes for an approximation of this form is equivalent to performing linear regressions with the sufficient statistics $T(x_{i}; x_{p_{i}})$ of the conditional approximations as explanatory variables and the (unnormalized) log posterior density as the dependent variable. Specifically, we show that the optimum in \eqref{eq:vbapprox} obeys the following fixed point condition:

\begin{eqnarray}
\label{eq:fixedpoint}
\eta_{i} & = & C_{i}^{-1}g_{i}, \text{ with},\\
C_{i} & = & \E_{q_{\eta}(x_{p_{i}})}\left\{\Var_{q_{\eta_{i}}(x_{i}|x_{p_{i}})}[T(x_{i};x_{p_{i}})]\right\} \\
g_{i} & = & \E_{q_{\eta}(x_{p_{i}})}\left\{\Cov_{q_{\eta}(x|x_{p_{i}})}[T(x_{i};x_{p_{i}}), \log p(x,y) - \log q_{\eta}(x) + \log q_{\eta_{i}}(x_{i}|x_{p_{i}})]\right\},
\end{eqnarray}

or, equivalently, that the \textit{natural gradient} of the KL-divergence is given by:

\begin{equation}
\label{eq:natgrad}
C_{i}^{-1} \nabla_{\eta_{i}} \D[q_{\eta}|p] = C_{i}^{-1}\E_{q_{\eta}(x_{p_{i}})}\left\{\Cov_{q_{\eta}(x|x_{p_{i}})}[T(x_{i};x_{p_{i}}), \log p(x,y) - \log q_{\eta}(x)]\right\}.
\end{equation}

Building on this result, we present an efficient stochastic approximation algorithm for solving the optimization problem in \eqref{eq:vbapprox}. In contrast to earlier work, our approach does not require any analytic calculation of integrals, which allows us to extend the fixed-form Variational Bayes approach to problems where it was previously not applicable.
\npar
Our method works by maintaining stochastic estimates of $C_{i}$ and $g_{i}$ as defined above and updating these during each iteration as

\begin{eqnarray}
C_{i,t+1} & = & (1-w)C_{i,t} + w\hat{\E}_{q_{\eta}(x_{p_{i}})}\left\{\Var_{q_{\eta_{i}}(x_{i}|x_{p_{i}})}[T(x_{i};x_{p_{i}})]\right\} \label{eq:cgupdate}\\
g_{i,t+1} & = & (1-w)g_{i,t} + w\hat{\E}_{q_{\eta}(x_{p_{i}})}\left\{\Cov_{q_{\eta}(x|x_{p_{i}})}[T(x_{i};x_{p_{i}}), \log p(x,y) - \log q_{\eta}(x) + \log q_{\eta_{i}}(x_{i}|x_{p_{i}})]\right\}, \nonumber
\end{eqnarray}

with a fixed stepsize $w$ and where the hat symbol ($\hat{\hspace{0.1cm}}$) denotes unbiased stochastic approximation. The parameters of $q_{\eta}(x)$ are then updated as

\begin{equation}
\eta_{i,t+1} = C_{i,t+1}^{-1}g_{i,t+1}.
\end{equation}

We find that approximating $C$ and $g$ using the same random draws eliminates most of the noise in our estimate of $\eta$, for the same reasons that the classical ordinary least squares estimator $(X'X)^{-1}X'y$ is statistically efficient for linear regression (and not theoretical alternatives like $\E[X'X]^{-1}X'y$). This noise cancellation is even stronger in our `regression' as our dependent variable $\log p(x,y)$ is \textit{noise free}, only the design points $x$ at which we calculate its value are random. Unlike with classical linear regression, using the same random draws to calculate both parts of the expression can introduce a slight bias into our estimates. By averaging our stochastic approximations over multiple iterations we eliminate this bias, make efficient use of all of our samples, and adapt our parameters $\eta$ slowly enough to ensure convergence of our algorithm.
\npar
In our earlier work \citep{linregvb} we left the implementation details to be decided by the user. Specifically, we assumed that the stepsize $w$, the number of iterations, the stochastic estimator, and the precise functional form of the approximation $q(x)$ were given. Here we give advice on how to make these decisions and we supply default choices. Our goal in defining these defaults is both to make it easier for other researchers to work with our methods as well as to make it possible to implement a fully automated version of our algorithm in statistical software packages like Stan \citep{stan-software:2013}. The main use-case we have in mind is to provide quick but accurate inference for hierarchical Bayesian models.

\section{Choosing the functional form of the approximation}
\label{sec:func_form}
Choosing an appropriate form of approximation $q(x)$ is essential to getting an accurate approximation to the posterior. In our earlier work we significantly expanded the choices that are available to the user, but we did not explicitly give advice on how to make this choice. We will try to do so here. For defining our default approximations we have in mind a possible future software package where a user is able to define a probabilistic model after which the software package will automatically output an appropriate form for the posterior approximation. Note that existing probabilistic programming packages like Infer.NET \citet{InferNET10} also work in this way. The main difference with our approach is how we choose to translate the model into a posterior approximation: rather than using the factorized approximations that are the default in earlier software packages we try to incorporate the dependencies of the model into the posterior approximation as much as possible in order to maximize the quality of the final approximation.  
\npar
We assume that the model and prior $p(x,y)=p(x)p(y|x)$ are given by a general probabilistic program that linearly runs over the unknows $x$ and randomly draws them one block at a time, after which the data $y$ is drawn last:

\begin{equation}
\label{eq:model}
p(x,y) = \left[\prod_{i} p(x_{i}|x_{p_{i}}) \right] p(y|x), 
\end{equation}

where $x_{p_{i}}$ once again denotes the `parents' of $x_{i}$, but now under the model. The data $y$ itself may also be generated in stages, like the unknowns $x$, but it is not necessary to explicitly consider this here: all that matters is that we are able to calculate the joint probability function $p(x,y)$ (possibly up to a normalizing constant) given all data, parameters, and latent variables in our model. Note that this setup does not allow for loops in the probabilistic program (unless they can be explicitly unrolled) or variable numbers of latent variables (as are common in applications of nonparametric Bayesian models), but that it is otherwise very general.
\npar
Our main strategy in automatically translating the model \eqref{eq:model} into a posterior approximation is simply this: try to have the posterior approximation match the model as closely as possible. This means we choose the approximation $q(x)$ to have the same hierarchical structure as the prior:

\begin{equation}
\label{eq:happrox}
q(x) = \prod_{i} q(x_{i}|x_{p_{i}}), 
\end{equation}

for the same partitioning $x_{1},\ldots,x_{i}$ of the parameters and latent variables. This obviously works well in capturing the dependencies of the model, but it also generally works well for capturing the posterior dependencies induced by the likelihood. The likelihood $p(y|x)$ is usually applied to the lowest level of the hierarchy, after which its effect works its way up the hierarchy through the dependencies of the model: if these dependencies are captured well, then generally so is the effect of the likelihood.
\npar
In choosing the functional form of the conditionals $q(x_{i}|x_{p_{i}})$ we will also try to match the prior as closely as possible. In the majority of cases, the conditional priors $p(x_{i}|x_{p_{i}})$ will already be in the exponential family:

\begin{equation}
\label{eq:priorexpfam}
\log p(x_{i}|x_{p_{i}}) = T(x_{i})\tilde{\eta}(x_{p_{i}}) - Z(x_{p_{i}}),
\end{equation}

where the natural parameters $\tilde{\eta}(x_{p_{i}})$ are a function of the parents of $x_{i}$. If the prior has this structure, we can simply re-use it as the conditional posterior approximation:

\begin{equation}
\label{eq:postexpfam}
\log q(x_{i}|x_{p_{i}}) = T(x_{i})\left[\tilde{\eta}(x_{p_{i}}) + \eta\right] - Z(\eta,x_{p_{i}}),
\end{equation}

where $\eta$ is a new set of variational parameters which will be used to capture the effect of the likelihood on the variable $x_{i}$. These new parameters are initialized to zero, or something weakly informative for improper priors, and will then be optimized by our stochastic optimization algorithm.
\npar
In case the conditional priors $p(x_{i}|x_{p_{i}})$ are not in the exponential family we will have to make some adjustments. The best solution is probably to define a number of default mappings that approximate specific non-exponential family distributions by specific appropriately chosen equivalents in the exponential family. For example, the scale of latent effects is often assigned a half-Cauchy prior \citep[e.g.][]{polson2012half}. It would make sense to replace this by an inverse-Gamma distribution on the squared scale in the approximation since that is the form of distribution that is often induced in the posterior. We could do the same for improper flat priors (as are often assigned to a location), which could be mapped to a Gaussian distribution. Another solution would be to re-express the non exponential family variables as transformations of multiple random variables in the exponential family. For example, a Cauchy distribution may be expressed as an inverse-Gamma scale mixture of Gaussians. The resulting inverse-Gamma and Gaussian variables are in the exponential family and their posterior can be approximated by a distribution of the same form.
\npar
The default form of approximation as defined here is very powerful, but in a way it does put the burden of defining a good approximation back on the user: it is often possible to specify (almost) the same model in different ways, and using the recipe described here these different specifications may not lead to the same form of posterior approximation. This is obvious for the specification of prior distributions, but more importantly the specification can also impact the structure of the chosen posterior approximation. As an example let us consider the stochastic volatility model discussed by \citet{riemannhmc}. The data we will use, from \citet{kim:98}, is the percentage change $y_{t}$ in the GB Pound v.s. US Dollar exchange rate, modeled as

\begin{equation}
\label{eq:proby}
y_{t} \sim N[0,\exp(v_{t})] \text{ for } t=1,2,\ldots,T,
\end{equation}

with volatilities $v_{t}$, governed by the autoregressive AR(1) process

\begin{equation}
\label{eq:probv}
v_{t+1} | v_{t} \sim N[\phi v_{t} + (1-\phi)\mu, \sigma^{2}], \text{ with } v_{1} \sim N[\mu,\sigma^{2}/(1-\phi^{2})].
\end{equation}

The priors of the parameters are given by

\begin{equation}
\label{eq:paramprior}
p(\mu) \propto c, \hspace{1cm} (\phi + 1)/2 \sim \text{Beta}(20,1.5), \hspace{1cm} \sigma^{2} \sim \text{Inv-Gamma}(5,0.25).
\end{equation}

\begin{figure}[H]
	\centering
		\includegraphics[width=\textwidth]{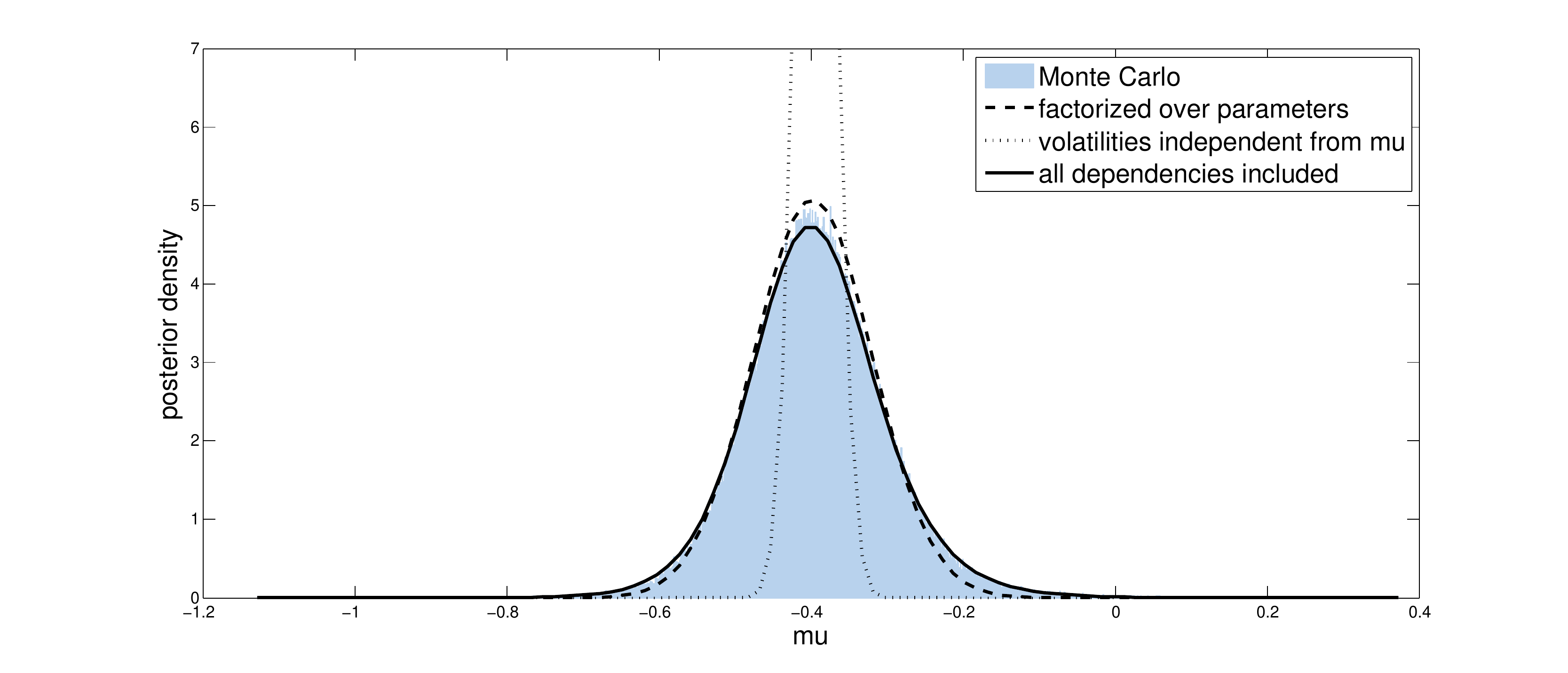}
	\caption{Posterior for the stochastic volatility model - $\mu$ parameter}
	\label{fig:svMu}
\end{figure}

\begin{figure}[H]
	\centering
		\includegraphics[width=\textwidth]{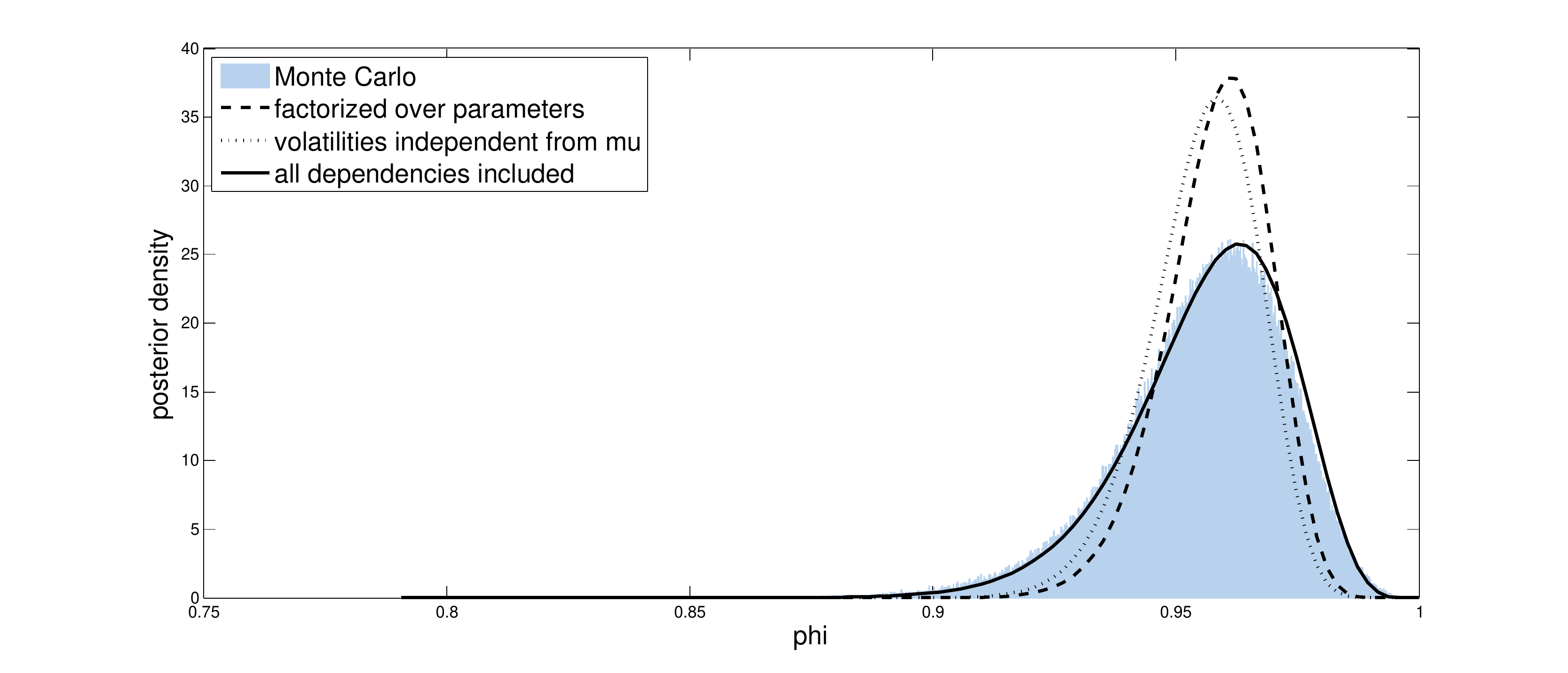}
	\caption{Posterior for the stochastic volatility model - $\phi$ parameter}
	\label{fig:svPhi}
\end{figure}

\begin{figure}[H]
	\centering
		\includegraphics[width=\textwidth]{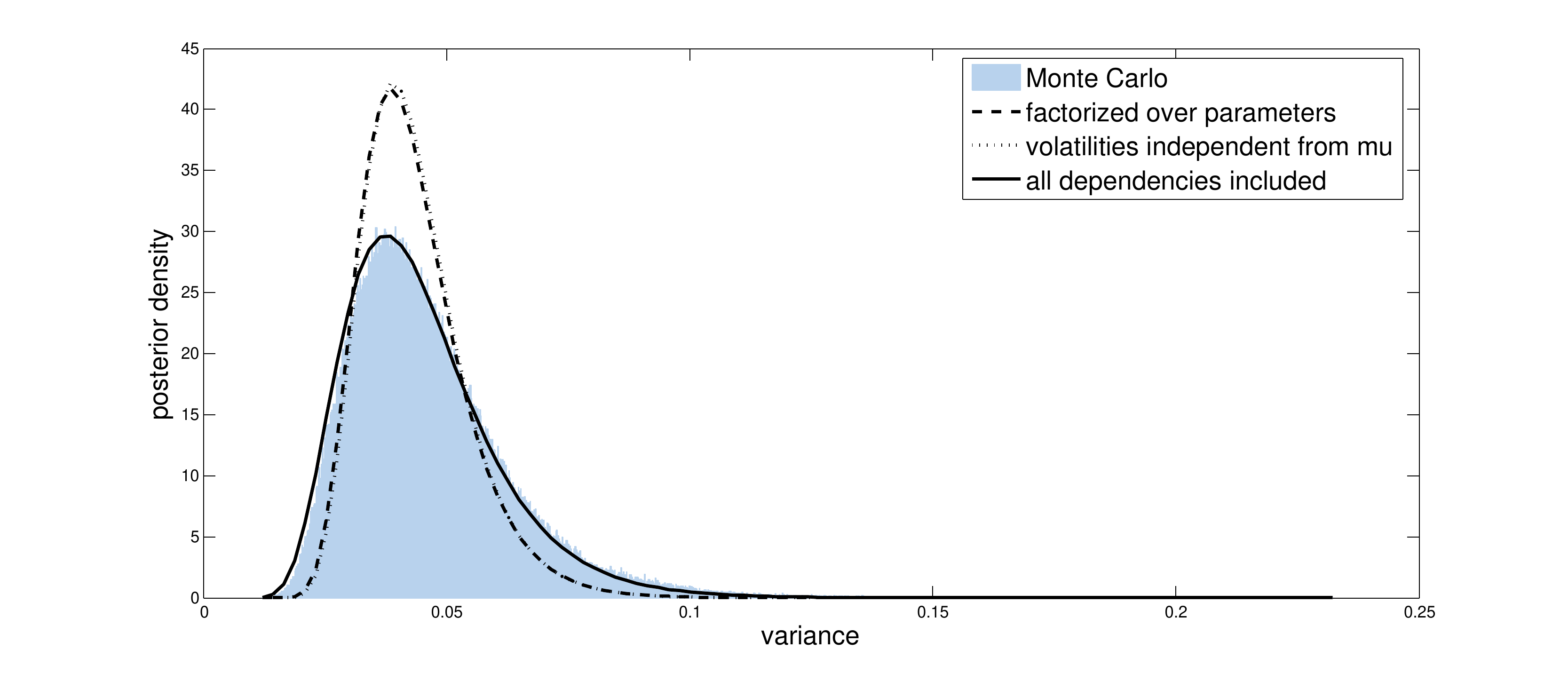}
	\caption{Posterior for the stochastic volatility model - $\sigma^{2}$ parameter}
	\label{fig:svVar}
\end{figure}

Applying our recipe for defining the posterior approximation to the model as specified above leads to an approximation that is factorized over the parameters $\mu,\phi,\sigma^{2}$ but that accurately incorporates the dependency of the volatilities $v$ on these parameters. As can be seen from Figures~\ref{fig:svMu}, \ref{fig:svPhi}, and \ref{fig:svVar}, this leads to a posterior approximation that slightly underestimates the posterior uncertainty in these variables, but that does a good job of capturing the means and general shapes of the marginal posteriors.
\npar
The model $p(\mu,\phi,\sigma^{2},y)$ as defined above could be specified equivalently using

\begin{equation}
y_{t} \sim N[0,\exp(\mu + v_{t})]
\end{equation}

in place of \eqref{eq:proby}, and using

\begin{equation}
v_{t+1} | v_{t} \sim N[\phi v_{t}, \sigma^{2}], \text{ with } v_{1} \sim N[0,\sigma^{2}/(1-\phi^{2})],
\end{equation}

in place of \eqref{eq:probv}. If we now turn our model specification into a posterior approximation, the result is an approximation in which the volatitilies $v$ are independent of the parameter $\mu$. As can be seen from Figures~\ref{fig:svMu}, \ref{fig:svPhi}, and \ref{fig:svVar}, this makes the posterior approximation worse: Although the approximation for $\sigma^{2}$ and $\phi$ do not change much, the posterior uncertainty in $\mu$ is now greatly underestimated. Moreover, the optimization now takes about 30 times as many iterations to converge, due to the volatilities $v$ and $\mu$ independently trying to model the same thing. However, do note that the approximation is still much better than what we would get using a posterior approximation that completely factorizes over the unknowns as is typical for applications of Variational Bayes: Although such an approach works well for some applications, it tends to give very misleading results for hierarchical models like the one considered here. 
\npar
By changing the way in which we specify our model, we can also improve the quality of the approximation. For this particular model we can view the joint prior $p(\mu,v|\phi,\sigma^{2})$ as the diffuse limit of a multivariate Gaussian, i.e.

\begin{equation}
\log p(\mu,v) = c[\phi,\sigma^{2}] + (\mu,v')\tilde{\eta}_{1}[\phi,\sigma^{2}] - \frac{1}{2}(\mu,v')\tilde{\eta}_{2}[\phi,\sigma^{2}](\mu,v')',
\end{equation}

with $c$ a univariate constant in $(\mu,v)$, $\tilde{\eta}_{1}$ a $T+1 \times 1$ vector and $\tilde{\eta}_{2}$ a sparse $T+1 \times T+1$ matrix representing the natural parameters of the diffuse joint prior $p(\mu,v|\phi,\sigma^{2})$. Writing the model in this way leads to a posterior approximation in which both $v$ and $\mu$ depend on the parameters $\phi,\sigma^{2}$. This specification might be a bit unnatural but a software package could implement this in a specialized function for defining dynamic models so that the user would not have to think about this. Note that software packages like Stan already incorporate specialized functions for defining dynamic models.
\npar
The only two unknowns that are now still independent in our posterior approximation are $\sigma^{2}$ and $\phi$, and we have to do something about this if we want a truly accurate approximation. Note that automatically detecting this independence in a software package is straightforward, but that fixing it automatically is a bit more difficult. One practical solution may be to simply incorporate a dependency on $\phi$ in the posterior approximation of $\sigma^{2}$ by setting

\begin{equation}
\log q_{\eta}(\sigma^{2}|\phi^{2}) = T(\sigma^{2})\left[\tilde{\eta} + \eta^{0} + \phi\eta^{1} + \phi^{2}\eta^{2}\right] - Z(\tilde{\eta},\phi,\eta^{0},\eta^{1},\eta^{2}), 
\end{equation}

where $T(\sigma^{2})$ represents a $1 \times 2$ vector containing the sufficient statistics of the inverse-Gamma distribution, $\tilde{\eta}$ are the parameters of the prior, and $\eta^{0},\eta^{1},\eta^{2}$ are separate $2 \times 1$ vectors of variational parameters. Adding a dependency of this form basically comes down to adding interaction terms to our regression. Indiscriminately adding such higher order terms runs the risk of making our optimization unstable, but also here the linear regression analogy comes to our rescue: by using a little bit of shrinkage on the coefficients $\eta^{1},\eta^{2}$ of the higher order terms we can regularize the solution of our variational optimization problem. We now have a posterior approximation with dependencies between all unknowns, which matches the true posterior distribution almost exactly as can be seen from Figures~\ref{fig:svMu}, \ref{fig:svPhi}, and \ref{fig:svVar}.
\npar
The approximation quality that users will get using our recipe for automatically specifying the approximation $q(x)$ will depend on the way in which they specify their model, but at least the general way in which to do this is clear: specify your model in such a way that you explicitly model as many of the dependencies between the different unknowns as possible. Ironically, this is precisely the opposite of what users of Stan have to do in order to get the MCMC sampler in that package to converge quickly. Possibly, a software package could make model specification easier for the user by automatically converting between different commonly occuring specifications of the same model. Futhermore, an important contribution of our earlier work \citep{linregvb} is the development of a measure of approximation quality, as discussed in Section~\ref{sec:quality}. Using this assessment of the approximation, a user of our hypothetical software package would at least know when a given specification leads to a poor approximation, and the user can experiment with different model specifications to see which gives the best approximation.

Due to convergence issues, fitting the posterior approximation with volatilities $v$ independent from $\mu$ takes between 10 and 20 seconds before a good quality approximation is found. Fitting the other two variational approximations took less than a second using our experimental MATLAB implementation.
\section{Checking the quality of the approximation}
\label{sec:quality}

In our earlier work \citep{linregvb} we make a connection between the optimization problem of fixed-form Variational Bayes \eqref{eq:vbapprox} and that of classical linear least squares regression: At its solution, the parameters of our posterior approximation are given by a linear regression using the unnormalized log posterior $\log p(x,y)$ as the dependent variable and the sufficient statistics of the posterior approximation as explanatory variables. A natural measure of the quality of our approximation that follows from this is the $R^{2}$ of that `regression':

\begin{eqnarray}
R^{2} = 1 - \frac{\Var_{q(x)}[\log p(x,y) - \log q_{\eta}(x)]}{\Var_{q(x)}[\log p(x,y)]}.
\end{eqnarray}

This measure of approximation quality can be calculated efficiently by drawing a small number of samples from the posterior approximation after convergence. In the context of variational approximation, the R-squared can be interpreted as the approximate remaining KL-divergence between our approximation and the true posterior, rescaled by the curvature in the true posterior.
\npar
We advice users of our variational approximations method to always use the R-squared to assess the accuracy of the approximation and to help select between different ways of specifying a model and its posterior approximation. A software implementation of our method should probably calculate this measure by default everytime an approximation is fitted.

\section{Choosing the stochastic estimator}
\label{sec:estim}
Stochastically approximating the two covariance terms $C$ and $g$ of our linear regression can be done in multiple ways. The simplest way is to draw samples from $q_{\eta}(x)$ and then to estimate $C$ and $g$ by the corresponding sample covariances, but in our earlier work we also introduced methods that could make use of the gradient of $\log p(x,y)$ and even its Hessian. In addition, we mentioned the possibility of `rotating' the linear system we are solving by instead approximating $\tilde{C}=K(\eta)C$ and $\tilde{g}=K(\eta)g$ using an invertible preconditioning matrix $K(\eta)$. Here we try to give guidance on how to choose the stochastic estimator and its preconditioner. We first discuss the preconditioning matrix and then deal with the different approximation methods in descending order of preference. In doing so, we focus on what estimator will be most efficient. In practice, ease of implementation will of course play a role as well, and the simple covariance estimator is more attractive from that perspective than when we only consider its efficiency.

\subsection{Preconditioning}
Our fixed point condition $\eta_{i} = C_{i}^{-1}g_{i}$ still yields the same result if we instead use $\tilde{C}_{i}=K(\eta_{i})C_{i}$ and $\tilde{g}_{i}=K(\eta_{i})g_{i}$ for an invertible preconditioning matrix $K(\eta_{i})$, since that matrix then cancels out when forming $\eta_{i}$. As we discuss in our earlier work, this allows us to estimate a `rotated' version of our regression that can sometimes be more efficient, but we did not yet specify which $K(\eta_{i})$ to use. After performing more experiments we can now say that for most applications the best preconditioner is given by

\begin{equation}
K(\eta_{i}) = \E_{q_{\eta}(x_{p_{i}})}\left\{\Var_{q_{\eta_{i}}(x_{i}|x_{p_{i}})}[T(x_{i};x_{p_{i}})]\right\}^{-1}.
\end{equation}

Using this preconditioner we have $\E_{q} \tilde{C}_{i} = \Id$, the identity matrix, which gives our linear system the best conditioning, and which most evenly weights the samples obtained over the different iterations of our optimization. Note that we can generally only calculate this $K(\eta_{i})$ for the top level of our hierarchical approximation, so this finding is mostly of theoretical interest in the current context, although we do observe some speed-up when using non-hierarchical approximations. For the type of hierarchical approximations considered here, we advice simply using the unrotated versions of our stochastic estimators.
\npar
As discussed in our earlier work, rotating the linear regression is equivalent to changing the parameterization in which we calculate the statistics of our regression. Using the preconditioner proposed here (in a non-hierarchical approximation) is then equivalent to using the \textit{mean parameterization} of our exponential family approximation, with parameters
\[
\mu(\eta) = \E_{q_{\eta}(x)} T(x).
\]
In our earlier work, we further show that when using an infinitesimal stepsize $w$ in combination with our unrotated estimators, our algorithm becomes equivalent to a preconditioned version of standard stochastic gradient descent. Using the preconditioner proposed here, a similar derivation shows that the algorithm becomes equivalent to a stochastic gradient descent procedure using unbiased estimates of the \textit{natural gradient}.
\npar
Finally, although it might be tempting to omit the stochastic approximation of $\tilde{C}_{i}$, since its expectation is $\Id$ for the given preconditioner, we find it is still very important to stochastically approximate this matrix in order to provide the noise cancellation that makes our method work.

\subsection{Analytic calculation}
The best stochastic estimator is one that is not stochastic. Often we are able to calculate some of the expectations in our variational approximation analytically, while others need to be approximated. For example, with the stochastic volatility model discussed in Section~\ref{sec:func_form} we are able to calculate analytically the expectation of the log likelihood with respect to the conditionally Gaussian approximation of the volatilities. Since this expectation can be calculated exactly, the volatilities do not need to be sampled. It would be useful if an automated version of our methods could detect situations like these, which may be made possible by using a software package for symbolic mathematics. An alternative would be to simply allow the user to program the analytical expectations directly into the model definition.

\subsection{Using the gradient and Hessian of the log posterior}
In our earlier work we show that
\begin{equation}
\label{eq:grad1}
C = \nabla_{\eta} \E_{q_{\eta}} T(x),
\end{equation}
and
\begin{equation}
\label{eq:grad2}
g = \nabla_{\eta} \E_{q_{\eta}} \log p(x,y),
\end{equation}
which suggests that there may be different, more efficient ways of approximating these statistics than just using the corresponding sample covariances. In particular, if (part of) our approximation is Gaussian, i.e. $q(x_{i}) = N(m,V)$, we can make use of the relationships
\begin{equation}
\nabla_{m} \E_{q}[h(x)] = \E_{q}[\nabla_{x}h(x)],
\label{op1}
\end{equation}
and
\begin{equation}
\nabla_{V} \E_{q}[h(x)] =  \frac{1}{2}\E_{q}[\nabla_{x}\nabla_{x}h(x)],
\label{op2}
\end{equation}
for any twice-differentiable function $h(x)$, as derived by \citet{MinkaThesis} and \citet{archambeau}, to approximate the regression statistics $C$ and $g$ using the gradient and Hessian of the log posterior and the sufficient statistics. This type of stochastic approximation typically has very low variance, and we recommend it for cases where the Gaussian distribution $q(x_{i})$ is either high-dimensional or when this distribution occurs at the bottom of our hierarchy. If $x_{i}$ is low dimensional and is used high up in the hierarchy we find it is usually more efficient to only use the first derivative (next subsection) and not the second.
\npar
In our earlier work we also derive a specialized efficient algorithm for Gaussian variational approximation based on a rotated version of our regression. It turns out that this algorithm is equivalent to what we obtain when using the preconditioning scheme proposed above. Finally, note that calculating the Hessian of the log posterior will also tell us which variables are conditionally independent: by taking into account these conditional independencies we can often set many of our variational parameters to zero beforehand which can save a lot of memory.

\subsection{Using the gradient and ignoring indirect effects}
If the variables $x_{i}$ in our model are continuous, we can also approximate \eqref{eq:grad1} and \eqref{eq:grad2} by differentiating the Monte Carlo estimator of $\E_{q} T(x)$ and $\E_{q} \log p(x,y)$. The result is an expression containing partial derivatives of the log posterior with respect to $x_{i}$, but that also includes terms that capture the indirect effect of $q_{\eta}(x_{i})$ on $\E_{q} \log q(x)-\log p(x,y)$ through its influence on the distribution of other variables $x_{j}$ down the hierarchy \citep[see][]{linregvb}. In our earlier work we already mentioned that we can usually ignore these indirect effects without causing a decrease in accuracy, since for most specifications the fixed points of the resulting expressions occur in exactly the same location. After doing more experiments, we can now not only confirm this, but we also find that ignoring these indirect effects often speeds up convergence of the algorithm. By ignoring the effect of $x_{i}$ on the distribution of $x_{j}$ down the hierarchy, we eliminate the `moving target' phenomenon that can otherwise cause the top level $q_{\eta}(x_{i})$ of our approximation to converge slowly when $q_{\eta}(x_{j}|x_{i})$ is still changing.

\subsection{Using the sample covariance}
The simplest, but least efficient, approximation method for estimating $C$ and $g$ is to simly use the corresponding sample covariances. When $x_{i}$ is discrete, this is currently the only estimator we have available. Note that we need to use multiple samples $x_{i}$ conditional on a single value of its parents $x_{p_{i}}$ when using a discrete distribution in a hierarchical approximation.
\section{Using a variable number of iterations}
\label{sec:nr_iter}

Our stochastic optimization algorithm proposed in \citet{linregvb} assumes we are given a fixed computational budget of $T$ iterations. We then set the stepsize $w$ used in \eqref{eq:cgupdate} equal to $1/\sqrt{T}$. In order to further reduce the noise in our final solution we also perform averaging over the last half of the iterations using

\begin{eqnarray}
\bar{C}_{i} & = & \sum_{t = T/2 + 1}^{T} \hat{C}_{i,t} \nonumber \\
\bar{g}_{i} & = & \sum_{t = T/2 + 1}^{T} \hat{g}_{i,t} \nonumber \\
\eta_{i,\text{final}} & = & \bar{C}_{i}^{-1}\bar{g}_{i}. \label{eq:avg}
\end{eqnarray}

This procedure is based on the work of \citet{robustsa} who show that such a procedure is optimal under certain conditions. In practice, setting the number of iterations $T$ beforehand is diffult since we typically do not have enough information to make a good time/accuracy trade-off. Here, we therefore propose to use a variable number of iterations in combination with a declining step-size

\begin{equation}
w_{t} = \frac{1}{\sqrt{10+t}},
\end{equation}

which offers the same asymptotic efficiency according to \citet{robustsa}. Compared to our original strategy this step-size causes us to take much larger steps in the beginning of our optimization procedure, which necessitates some changes in how we update our estimate of the parameters $\eta$. Instead of setting $\eta_{i,t} = C_{i,t}^{-1}g_{i,t}$ at every iteration, we use the damped update

\begin{eqnarray}
\eta_{i,t} & = & \alpha_{t} C_{i,t}^{-1}g_{i,t} + (1-\alpha_{t})\eta_{i,t-1}, \text{ with } \label{eq:damped} \\
\alpha_{t} & = & \min\left[1, \sqrt{\frac{c_{\text{mss}} K}{\sum_{i} d_{i,t}'d_{i,t}}}\right], \text{ and } \\
d_{i,t} & = & C_{i,t}^{-1}g_{i,t} - \eta_{i,t-1},
\end{eqnarray}

where $K$ is the total number of parameters in our variational approximation and $c_{\text{mss}}$ is a constant determining our maximum mean-square step size. We find that the precise value of $c_{\text{mss}}$ does not matter much and we fix it to 10 in all of our experiments. In addition to limiting the step size in this way we also check before every update that the proposed new $\eta$ value defines a proper distribution (and not, say, a negative variance). If this is not the case we further decrease $\alpha_{t}$ for that iteration. We find that this is enough to keep the algorithm under control early on during the optimization. After a few iterations the algorithm becomes more stable and we typically have $\alpha_{t} = 1$ at every subsequent iteration. Note that this approach delays the effect of the stochastic estimates $\hat{C}_{i,t},\hat{g}_{i,t}$ if $\alpha_{t} < 1$ but that it does not permanently decrease it as would happen if we would reduce the step size $w_{t}$ directly. Adjusting $w_{t}$ based on the stochastic iterates runs the risk of biasing our estimates and is therefore not to be recommended.
\npar
We assess convergence by checking whether $z_{t} \leq c_{\text{tol}}$, where $z_{t}$ is a running average of 
\[
\frac{1}{K} \sum_{i} d_{i,t}'d_{i,t},
\]
and where $c_{\text{tol}}$ is a pre-specified tolerance value depending on the degree of accuracy required. After this convergence check is met we continue for a set number of iterations $N$ (say 50), and do the averaging of \eqref{eq:avg} on these last iterations after which we output the final solution.

\section{Alternative: Batch approximation with line-search}
\label{sec:linesearch}
The stochastic optimization procedure discussed in Section~\ref{sec:nr_iter} is highly efficient in the way it uses our samples, but may not be very easy to implement efficiently and robustly in a general software package. An easier approach may be to use the stochastic estimators discussed in Section~\ref{sec:estim} to estimate the natural gradient for a fixed random seed and fixed (larger) number of samples and to use this gradient estimate in a standard line-search based optimization procedure. Algorithm~\ref{algo:batchvb} shows how such a gradient would be calculated.

\begin{algorithm}[H]
\caption{Batch approximation of the natural gradient of the KL-divergence}
\label{algo:batchvb}
\begin{algorithmic}

\REQUIRE A fixed random number seed $s^{*}$, e.g. $s^{*} = 12345$ 
\REQUIRE The total number of samples $N$
\REQUIRE The current value of $\eta$

\STATE Set the random number seed to the given number, i.e. $s \leftarrow s^{*}$
\STATE Draw N samples from the approximate posterior, $x^{*} \sim q_{\eta}(x)$
\STATE Use $x^{*}$ to approximate the natural gradients w.r.t. each $\eta_{i}$ using \eqref{eq:natgrad} and Section~\ref{sec:estim}
\RETURN $f$ = the combined vector of estimated natural gradients

\end{algorithmic}
\end{algorithm}

If our model only contains continuous variables, the multivariate function $f(\eta)$ calculated by Algorithm~\ref{algo:batchvb} is smooth in the parameters $\eta$, and we can find a solution to $f(\eta)=0$ using standard gradient-based optimization methods. In particular, we have performed successful experiments using natural gradient descent with updates of the form

\begin{eqnarray}
\eta_{t+1} & \leftarrow & \eta_{t} + \alpha s \text{ with}, \\
s & = & -f(\eta_{t}),
\end{eqnarray}

where we search for an acceptable step-length $\alpha \in (0,\alpha_{\text{max}}]$ using quadratic interpolation of $f(\eta)$ along the search direction $s$. Here we initialize $\alpha_{\text{max}}$ to 
\[
\alpha_{\text{max}} = \min\left[1.5, \sqrt{\frac{K c_{\text{mss}}}{s's}}\right],
\]
with $K$ once again the number of parameters in $\eta$ and with $c_{\text{mss}}$ a maximum mean-squared step size which we can set to something large like 100 or 1000. At every line search we start by proposing $\alpha = \min(1,\alpha_{\text{max}})$ which is the natural choice for any quasi-Newton optimization procedure. Subsequently, we (temporarily) reduce $\alpha_{\text{max}}$ whenever we encounter a value of $\eta$ that does not define a proper distribution $q_{\eta}(x)$.
\npar
Determining whether a given $\alpha$ is an acceptable step-length is not completely straightforward as we can only calculate the natural gradient $f(\eta)$ of the KL-divergence, and not the KL-divergence itself. Note that our sample estimate of the natural gradient does not correspond to a sample estimate of the KL-divergence: we find that sample estimates of the KL-divergence are extremely noisy, while the estimate of its natural gradient is often highly precise due to the noise cancellation of our stochastic linear regression. We have had success accepting a step length $\alpha$ if either
\[
\alpha = \alpha_{\text{max}} \text{ and } s'f(\eta_{t} + \alpha s) < 0,
\]
or
\[
|s'f(\eta_{t} + \alpha s)| < c |s'f(\eta_{t})|,
\]
with $c$ a constant that we set to $0.5$ in our experiments. This last condition is known as the \textit{strong curvature condition}. More sophisticated gradient-only line search strategies than the one outlined here may exist in the literature, but we do not expect a huge improvement from them: early on in the optimization we usually find that the first few proposals are either accepted directly or rejected for producing an improper $q_{\eta}(x)$, while later on the initial proposal with $\alpha=1$ is often accepted directly. Following \citet{honkela}, we also experimented with using search directions conjugate to the natural gradient, but we found this to only slow down the optimization. The reason for this might be that we calculate the natural gradient in the natural parameterization of our posterior approximation, which we generally find to give a better search direction than what we get when using the different parameterizations used by these researchers. Another reason may be that our posterior approximations often incorporate more dependencies between the different variables compared to typical applications of variational Bayes, which further improves the convergence when following the natural gradient. 
\npar
Unlike the algorithm in Section~\ref{sec:nr_iter}, the batch version of our optimization algorithm needs more than a single sample per iteration. The gradient-based estimators of Section~\ref{sec:estim} really are very efficient though, so we do not need very many. For the stochastic volatility model discussed in Section~\ref{sec:func_form} we obtain good results using as few as 5 samples. For models with continuous variables we can straightforwardly calculate the minimum number $n_{\text{min}}$ of samples required for all $C_{i}$ matrices to be invertible, which we can then multiply by a small constant to determine the number of samples we should use. A particularly efficient strategy is to first do the optimization for a small number of samples, say $2 n_{\text{min}}$, and then finetune the solution using a somewhat larger number, like $5 n_{\text{min}}$.
\npar
The increased number of samples per iteration compared to Section~\ref{sec:nr_iter} is partly offset by the fact that the batch version typically needs fewer iterations to converge. In addition, the multiple samples per iteration can easily be processed in parallel on a multicore CPU, or perhaps even on a graphics card for large problems, while the algorithm from Section~\ref{sec:nr_iter} benefits less from such parallelization. All in all, we find that the batch version of our algorithm is typically about equally fast as our online algorithm.
\npar
While our batch optimization procedure does not suffer from a lack of speed, it does have the serious disadvantage that it is not well suited for models with discrete variables. Discrete distributions introduce discontinuities into our estimated natural gradient $f(\eta)$ which can cause trouble with the line-search procedure. Also, they generally require a larger number of samples. Note that even continuous variables can cause discontinuities if they are sampled using a procedure involving an acceptance/rejection step, as is typical for distributions like the Gamma. However, in our experiments we find such discontinuities to be small enough not to cause problems. Another disadvantage of our batch method is that is not suited to work on streaming data.
\section{Conclusion}
\label{sec:conclusion}

We think the stochastic approximation method proposed in \citet{linregvb} is very powerful: By making use of the noise cancelling properties of a linear regression, it enables us to estimate the natural gradient of the KL-divergence between an approximation $q_{\eta}(x)$ and its target $p(x|y)$ at very low computational cost for a very large class of models and approximations. By combining this with an efficient optimization algorithm it opens up a lot of new applications to variational Bayes. The hierarchical posterior approximations it enables are especially promising for performing fast but accurate inference in hierarchical models where classical factorized approximations tend to do poorly. Nevertheless, implementing our `stochastic linear regression' variational approximations is not easy: there are lots of different choices to be made and all are important if we are to get the most out of this method. In this document we have offered our advice on how to make these choices. It is our hope that the tips and defaults defined here will help shed some light on how best to use this method in practice, and perhaps enable a fully automated implementation in the future.

\clearpage
\bibliographystyle{ba}
\bibliography{biball}

\end{document}